\documentclass[10pt,conference]{IEEEtran}
\IEEEoverridecommandlockouts

\usepackage{booktabs} 
\usepackage{listings}
\usepackage{graphicx}
\usepackage{amsmath}
\usepackage{color}
\usepackage{algorithm}
\usepackage[noend]{algpseudocode}

\algrenewcommand\algorithmicindent{0.7em}
\usepackage[flushleft]{threeparttable}
\usepackage{multirow}
\usepackage{romannum}
\usepackage{balance}
\usepackage{enumitem}
\usepackage{cite}
\usepackage{cleveref}

\def\BibTeX{{\rm B\kern-.05em{\sc i\kern-.025em b}\kern-.08em
    T\kern-.1667em\lower.7ex\hbox{E}\kern-.125emX}}

\begin{document}
\title{Revisiting ssFix for Better Program Repair}

\author{\IEEEauthorblockN{Qi Xin and Steven P. Reiss}
\IEEEauthorblockA{Department of Computer Science\\
Brown University\\
Providence, RI, USA\\
\{qx5,spr\}@cs.brown.edu}}

\maketitle

\begin{abstract}
A branch of automated program repair (APR) techniques look at finding
and reusing existing code for bug repair. ssFix is one of such techniques
that is syntactic search-based: it searches a code database for code
fragments that are syntactically similar to the bug context and reuses 
such retrieved code fragments to produce patches. Using such a syntactic 
approach, ssFix is relatively lightweight and was shown to outperform 
many other APR techniques. In this paper, to investigate the true 
effectiveness of ssFix, we conducted multiple experiments to validate
ssFix's built-upon assumption (i.e., to see whether it is 
often possible to reuse existing code for bug repair) and evaluate its 
code search and code reuse approaches. Our results show that while the 
basic idea of ssFix, i.e., reusing existing code for bug repair, is promising,
the approaches ssFix uses are not the best and can be significantly improved. 
We proposed a new repair technique sharpFix which follows ssFix's basic idea 
but differs in the code search and reuse approaches used. We evaluated 
sharpFix and ssFix on two bug datasets:
Defects4J and Bugs.jar-ELIXIR. The results confirm that sharpFix is an
improvement over ssFix. For the Defects4J dataset, sharpFix successfully repaired 
a total of 36 bugs and outperformed many existing repair techniques in
repairing more bugs. For the Bugs.jar-ELIXIR dataset, we compared sharpFix, 
ssFix, and four other APR techniques, and found that sharpFix has the best 
repair performance. In essence, the paper shows how effective a syntactic 
search-based approach can be and what techniques should be used for such an 
approach.
\end{abstract}


\section{Introduction}

Automated program repair (APR) \cite{monperrus18automatic} can significantly
save people time and effort by repairing a bug\footnote{In this paper, 
we use ``bug'' and ``fault'' interchangably.} automatically. Given a faulty 
program and a fault-exposing test suite, a typical APR technique automatically 
modifies the faulty program to produce a patched program that passes the 
test suite. Over the past decade, many APR techniques
\cite{legoues12,weimer13,kim13,y_qi14,ke15,z_qi15,xuan16,le16,long16,mechtaev16,
le17s3,xiong17,chen17contract,saha17elixir,xin17leveraging,liu18mining,
jiang18shaping,hua18towards,mechtaev18semantic,wen18context} have been developed. 
They look at using different approaches to do bug repair. One major problem 
faced by current APR techniques is the search space problem
\cite{long16-anal}.  An APR technique often needs to define a huge search space 
of patches to support repairing different types of bugs. However, searching for 
a correct patch in such a huge search space is often difficult \cite{long16-anal}.

To address the problem, one idea is to reuse existing code from existing programs. 
By doing so, a repair technique can avoid generating a large amount of artificial 
code to possibly avoid the search space explosion. The recent APR technique ssFix
\cite{xin17leveraging} was built upon the idea. It performs syntactic code search 
to find existing code fragments
(from the local faulty program and an external code repository) that are similar
to the bug context and reuse those code fragments to produce patches for bug repair. 
ssFix leverages the syntactic differences between any retrieved code fragment and 
the bug context to produce patches. For a code fragment that is similar to the bug 
context, the differences are small, and the search space is therefore reduced. 
Experimental results from \cite{xin17leveraging} showed that ssFix
is lightweight and relatively effective: it generated valid patches for 20 bugs in 
the Defects4J dataset \cite{defects4j} with the median running time for generating 
a patch being only about 11 minutes, and it outperformed five other APR techniques
\cite{legoues12,z_qi15,xuan16,le16,xiong17}.

Though ssFix seems like a promising technique, we do not know much about
its repair potential for two reasons: First, ssFix is built upon the assumption
that existing programs contain the fix ingredients (the statements/expressions 
needed for producing a correct patch). However, we do not really know how often 
the assumption holds in practice. Second, assuming the fix ingredients do exist 
in existing programs, we do not know whether ssFix uses the best approaches 
for finding and reusing these fix ingredients for repair.

In this paper, we conducted multiple experiments on the Defects4J bug dataset
\cite{defects4j} to (1) test the fix-ingredient-assumption, i.e., to investigate
whether the fix ingredients for bug repair often exist and to (2) evaluate 
ssFix's abilities in finding and reusing the fix ingredients for bug repair. For (1),
we first defined the fix ingredient for a bug in the context of automated program 
repair (\Cref{def_fi}) and then performed syntactic code search to check whether 
the fix ingredient exists in the local faulty program or in any non-local programs 
in a code repository (\Cref{test_fi_expt}). For (2), we conducted multiple 
experiments (\Cref{eval_ssfix_compt}) looking at (a) whether ssFix can effectively 
retrieve the existing fix ingredients, (b) whether ssFix can effectively reuse the 
fix ingredients it retrieved to do bug repair, and (c) whether ssFix can do effective 
repair possibly with the fault being accurately located.

Our results showed that (1) the idea of reusing existing code for bug repair is
promising (\Cref{test_fi_expt_rslt}) and (2) the approaches used by ssFix for code
search and code reuse however can be improved significantly
(\Cref{eval_ssfix_codesearch,eval_ssfix_codereuse}).
Based on our experimental observations, we developed a new repair technique
sharpFix which follows ssFix's basic idea but uses improved approaches for
code search and reuse. sharpFix improves ssFix's code search by using different
code search methods for retrieving code fragments from the local faulty
program and from the non-local programs in the code repository. For patch generation, 
sharpFix goes through the same steps used by ssFix: code translation, 
code matching (component matching in \cite{xin17leveraging}), and modification.
Each step however is different and improved. The improved approaches used by
sharpFix let it correctly repair 36 Defects4J bugs in total and outperform 
ssFix in correctly repairing 14 more bugs. We further evaluated sharpFix,
ssFix, and four existing APR techniques: jGenProg \cite{astorimpl},
jKali \cite{astorimpl}, Nopol (version 2015) \cite{nopolimpl}, and 
HDRepair \cite{hdrepairimpl} on another dataset: Bugs.jar-ELIXIR \cite{saha17elixir}
which contains 127 real Java bugs. The results show that sharpFix outperformed 
all these techniques and again confirm sharpFix is an improvement over ssFix.

This paper essentially shows how effective a syntactic search-based approach
can be and proposes an effective technique that should be used for such an 
approach. It makes the following contributions:
\begin{itemize}[noitemsep]
\item We show through experiments that the idea of reusing existing code for
repair is promising but ssFix's current approaches for code search and 
reuse can be significantly improved.
\item We developed a repair technique sharpFix (code is available at
\textit{https://github.com/sharpFix18/sharpFix}) which follows ssFix's 
basic idea but uses improved approaches for code search and code reuse.
\item We evaluated sharpFix and ssFix on two bug datasets: Defects4J 
and Bugs.jar-ELIXIR. We also evaluated four existing APR techniques: 
jGenProg, jKali, Nopol, and HDRepair on the latter dataset. The results 
showed that sharpFix outperformed all these techniques.
\end{itemize}

\section{Testing the Fix-Ingredient-Assumption} \label{test_fia}

A search-based repair technique like ssFix does bug repair through reusing
existing code and is built upon an assumption that existing code that
it looks at often contains the fix ingredient needed for producing a correct
patch. It is important to know whether assumption holds in practice to 
understand whether a search-based repair technique is truly effective. We 
conducted an experiment to test the assumption, i.e., to investigate how
often the fix ingredient for bug repair exists. For our experiment, we looked
at simple patches. For a simple patch, all the fixing changes are made within
either an expression or a primitive statement which contains no children statements.
In the context of automated bug repair, we defined six types of fix ingredients
for a simple patch. For each bug in the Defects4J dataset whose patch is simple,
we identified the fix ingredient and checked whether it exists in the code 
database we used. We did not look at more complex patches. To deal with a complex
patch, one may apply our method by first dividing the patch into simple patches 
and then searching for the fix ingredient for each. This corresponds to a
natural way of producing a complex patch: it is not likely to produce the
component (simple) patches all at once but one by one.

There are existing studies that also test the fix ingredient assumption:
Some of the studies \cite{barr14,sumi15} looked at whether the fix ingredient
exists at the level of code lines. A fix ingredient however does not have to 
be an entire code line to be reused for bug repair (we will show this with an
example later). The others \cite{nguyen13study,martinez14} do not actually fit
our context. For example, Martinez et al. \cite{martinez14} studied whether 
the fix ingredient exists in a program's previous versions which a repair 
technique like ssFix often does not look at for repair.

\subsection{Defining the Fix Ingredient} \label{def_fi}

Some simple definitions can be either too general or too constrained. 
As an example, the correct (developer) patch for the Defects4J
bug \textit{Math\_85} changes the buggy if-condition 
from \texttt{fa*fb>=0.0} to \texttt{fa*fb>0.0}.
A fix ingredient defined as the exact changed expression \texttt{>} is too
general: we can find lots of code fragments containing \texttt{>}, but very
few of them can possibly be useful for repair. A fix ingredient defined as the 
changed line can be constrained since it requires the correct expression
to appear as an if-condition. A fix ingredient defined as the changed statement
is too program-specific and thus too constrained.

In the context of automated bug repair, we looked at using six types of modifications
(M0-M5 as shown below) used by existing APR techniques 
\cite{legoues12,long15,long16,xiong17,xin17leveraging} to model a general repair
modification. For each type of modification, we defined the corresponding fix 
ingredient. If the fix ingredient can be found within a code fragment that is 
reasonably small, then it is possible for a repair technique to effectively reuse 
the fix ingredient to produce a correct patch.
\begin{itemize}
\item \textbf{M0}: Combining a Boolean condition with another Boolean condition
using \texttt{\&\&} or \texttt{||} (e.g., \texttt{if (c0) \{...\}} $\rightarrow$
\texttt{if (c0 || c1) \{...\}})
\item \textbf{M1}: Changing an expression (as a non-if-condition) to another
expression (also as a non-if-condition) (e.g., \texttt{e0} $\rightarrow$ \texttt{e1})
\item \textbf{M2}: Changing an if-condition to another if-condition
(e.g., \texttt{if (c0) \{...\}} $\rightarrow$ \texttt{if (c1) \{...\}})
\item \textbf{M3}: Adding an if-condition for one or more statements
(e.g., \texttt{s} $\rightarrow$ \texttt{if (c) \{s\}})
\item \textbf{M4}: Replacing a statement with another statement
(e.g., \texttt{s0} $\rightarrow$ \texttt{s1})
\item \textbf{M5}: Inserting a statement
(e.g., \texttt{s} $\rightarrow$ \texttt{s0; s})
\end{itemize}
The corresponding fix ingredients (FIs) are shown below.
\begin{itemize}
\item \textbf{FI0}: The combined Boolean condition (\texttt{c1} in \textbf{M0}'s example)
\item \textbf{FI1}: The parent statement/expression of the changed expression 
(the parent of \texttt{e1} in \textbf{M1}'s example)
\item \textbf{FI2}: The changed if-condition (\texttt{c1} in \textbf{M2}'s example)
\item \textbf{FI3}: The added if-condition (\texttt{c} in \textbf{M3}'s example)
\item \textbf{FI4}: The replaced statement (\texttt{s1} in \textbf{M4}'s example)
\item \textbf{FI5}: The inserted statement (\texttt{s0} in \textbf{M5}'s example)
\end{itemize}
For M0, M2, and M3, we use the combined, changed, and added if-conditions as 
the fix ingredients respectively. For M4 and M5, we use the replaced and inserted 
statements as the fix ingredients respectively. For M1, we do not
simply use the changed expression (i.e, \texttt{e1} in \textbf{M1}'s example)
as the fix ingredient since the changed expression can be too small
and thus be lack of context (consider \texttt{e1} as a variable argument of a 
method call). So instead, we use the parent statement/expression of the changed 
expression as the fix ingredient (in the abstract syntax tree, this is the parent 
node of the changed expression node).

For the \textit{Math\_85} example, we define the fix ingredient as the
expression \texttt{fa*fb>0.0}, and we successfully found the exact fix 
ingredient from the local faulty program in a loop statement:
\texttt{do \{...\} while((fa*fb>0.0)\&\&...)}. For this bug, this is the 
only statement in the local program where \texttt{fa*fb>0.0} is contained.

\subsection{Experiment for Testing the Fix Ingredient Assumption} \label{test_fi_expt}

\subsubsection{Setup}

Our experimental dataset is the Defects4J dataset (version 0.1.0) \cite{defects4j}
which contains in total 357 real bugs. We manually
identified 103 bugs whose developer patches (provided by the dataset) are
simple. For each patch, we looked at the modification types in an sequential 
order from M0 to M5, identified the first type as which the patch can be 
classified, and then identified the fix ingredient. For experiment, we checked
whether the fix ingredient exists in either the local faulty program or any 
non-local program in a code repository for which we used the DARPA MUSE code 
repository \cite{darpamuse} which consists of 66,341 Java projects (about 81GB).

\subsubsection{Search Methodology}

For each simple patch, we performed syntactic code search to check whether 
the identified fix ingredient can be found within the code database, i.e., 
the local faulty program plus all the programs in the code repository we used.
In our definition, a fix ingredient can only be an expression or a statement.
So we extracted every statement within every method in the code database, 
tokenized the fix ingredient and every extracted statement, and checked whether
the fix ingredient's tokens are a subsequence of any extracted statement's
tokens in either the exact or the parameterized form. For parameterization,
we replaced program-specific (i.e., non-JDK) variables, types, and methods 
with special symbols: \texttt{\$v\$} for variables, \texttt{\$t\$} for types, 
and \texttt{\$m\$} for methods.

To find a fix ingredient within the local faulty program, we exactly checked
whether the fix ingredient's tokens are a subsequence of any extracted statement's
tokens with and without parameterization. To find a fix ingredient within the code
repository, we could do the same, but this can be very expensive: for each of the
103 bugs, we would have to iterate every statement in the large code repository.
So instead, we indexed every statement in the code repository using ssFix 
(from \textit{https://github.com/qixin5/ssFix}), and performed
two steps: (1) we did ssFix's code search using the fix ingredient's enclosing 
statement as the query to have the top-500 statements retrieved and (2) we checked, 
for each retrieved statement, whether the fix ingredient's tokens are a subsequence 
of the statement's tokens. For code search within the repository, we filtered
away any statement whose enclosing method's signature is identical to the faulty
method's signature and whose enclosing package's name is identical to the faulty
method's enclosing package's name (as did in \cite{xin17leveraging}). We did so
to ignore any fix ingredients that are simply from any bug-fixed versions of the
faulty program.

\subsubsection{Results} \label{test_fi_expt_rslt}

\begin{figure}
\centering
\includegraphics[width=0.35\textwidth]{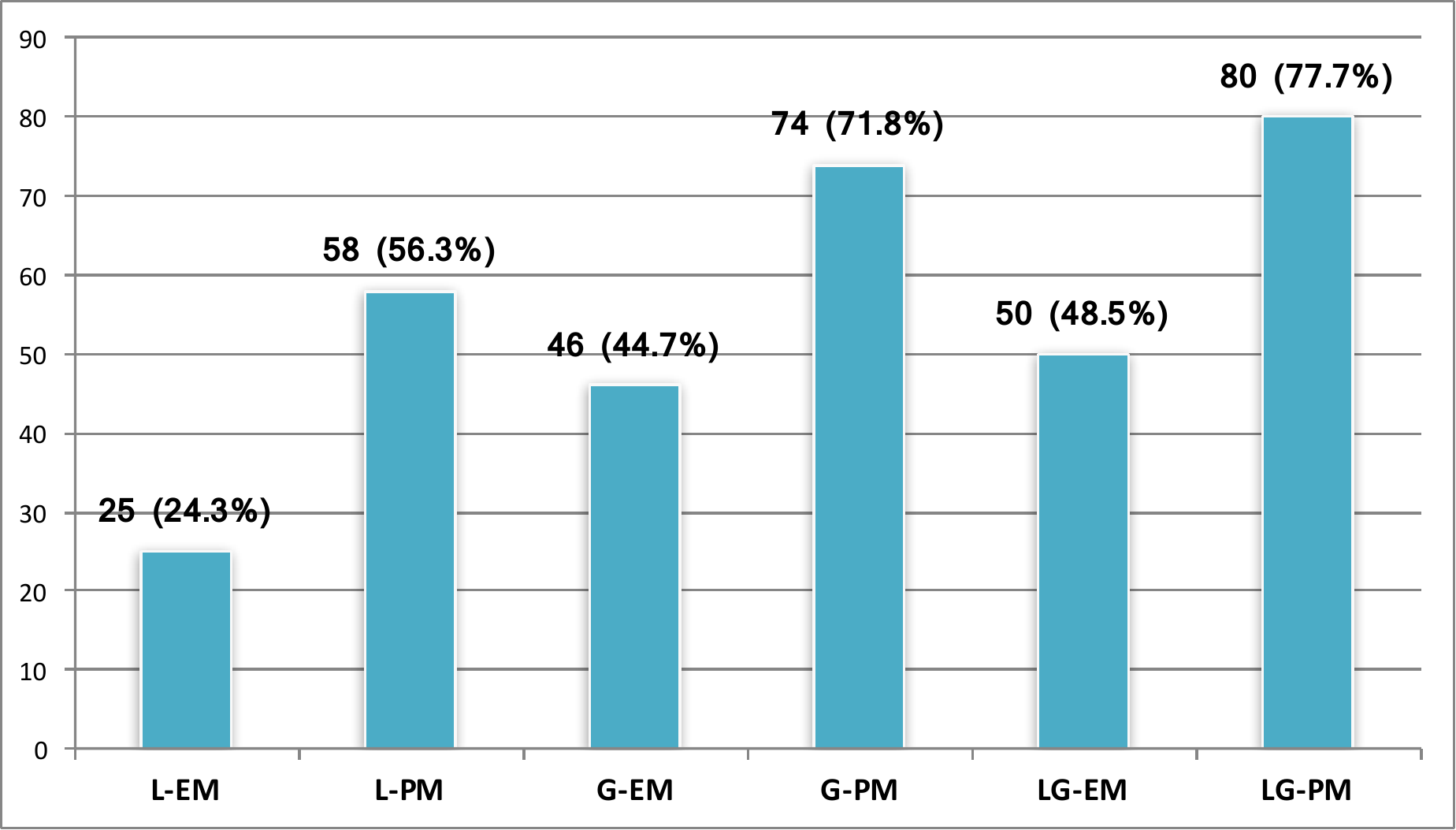}
\caption{Fix Ingredient Retrieval Result (L/G/LG-EM: Exact Match within the
local program/code repository/both; L/G/LG-PM: Parameterized Match within
the local program/code repository/both. The column shows the numbers and the 
percentage of bugs for which the fix ingredients were found)}
\label{fi_search_rslt0}
\vspace{-1.5em}
\end{figure}

\Cref{fi_search_rslt0} shows our results.
We found that reusing existing code for bug repair is a promising approach:
For 50 (48.5\%) of the 103 bugs, we retrieved
the exact fix ingredients from the local faulty program and the code repository
(i.e., LG-EM). If the fix ingredient can be found within a code fragment that is 
reasonably small, a repair technique can possibly leverage the fix ingredient to 
produce a correct patch. For a total of 80 (77.7\%) of the 103 bugs, we retrieved 
fix ingredients in the parameterized forms (i.e., LG-PM). For 30 (80-50) bugs, we 
retrieved fix ingredients that are \textit{only} in the parameterized forms. 
Note that it is possible to correctly repair a bug without finding the fix ingredient 
we defined, so our results only provide a lower-bound.

\section{Analyzing ssFix} \label{eval_ssfix_compt}

Given that the fix ingredients often exist, the next question is whether a 
search-based repair technique can effectively find and reuse them for repair.
We analyzed ssFix by conducting experiments to evaluate its code search, code
reuse, and repair abilities. In this section, we show the experiments we 
conducted and the results we got. We found that the current code search and 
the code reuse approaches used by ssFix can be significantly improved.

\subsection{Some Background of ssFix}

Given a faulty program, a fault-exposing test suite (which the program failed), and
a code database which consists of the faulty program and a large code repository,
ssFix goes through the following four stages to possibly produce a patched program 
that can pass the test suite. Such a patched program (or the patch) is called
\textit{plausible}.
\begin{enumerate}
\item \textbf{Fault Localization}: ssFix relies on an existing technique GZoltar 
\cite{campos12} to identify a list of suspicious statements 
ranked from the most suspicious to the least. In the following stages, it looks 
at each statement independently to produce patches.
\item \textbf{Code Search}: Given a suspicious statement, ssFix produces a code chunk
(called the \textit{target} chunk, or the \textit{target}) including the statement 
possibly with its local context and searches for code chunks (called the 
\textit{candidate} chunks, or the \textit{candidate}s) in the code database that are
syntax-related (structurally similar and conceptually related) to the target to be 
reused for bug repair. In later stages, ssFix looks at reusing each candidate to 
produce patches for the target independently.
\item \textbf{Patch Generation}: Given a candidate being retrieved, ssFix first
translates the candidate by renaming the variables, types, and methods used in the
candidate and then produces a set of patches based on the syntactic differences between
the target and the translated candidate. More specifically, ssFix matches the related
statements and expressions between the two chunks and performs three types of 
modifications to produce patches based on the matching result.
\item \textbf{Patch Validation}: ssFix sorts the generated patches and then tests
them to find the first plausible patch. For testing a patch, ssFix first
applies the patch to the faulty program to get a patched program. It then checks
whether the patched program can compile and can pass the test suite.
\end{enumerate}
More details about ssFix can be found in \cite{xin17leveraging}. In this paper,
we call the last two stages \textit{code reuse}.

\subsection{Evaluating ssFix's Code Search} \label{eval_ssfix_codesearch}

To evaluate ssFix's code search ability, we looked at the 103 bugs used 
in \Cref{test_fi_expt} whose developer patches are simple. For each bug,
we provided ssFix with the real faulty statement, ran its code search, 
and checked whether it can effectively retrieve any candidates that contain 
the fix ingredient that we identified earlier. We call a candidate
(possibly after translation) that contains the fix ingredient in its exact 
form \textit{\textbf{promising}}. Our results show that ssFix retrieved 
promising candidates within the top-500 results for 38 bugs.

\subsubsection{Experiment} \label{eval_ssfix_codesearch_expt}

For each of the 103 bugs, we provided ssFix with the faulty statement,
ran ssFix's code search to retrieve a list of ranked candidates
(as code chunks) from the code database, translated the candidates using ssFix's
code translation (otherwise it may not be able to reuse the fix ingredient 
for repair), and checked whether any retrieved candidate is promising, i.e., 
contains the exact fix ingredient\footnote{For a bug whose patch inserts a 
statement in between two contiguous statements, we used each of the two statements 
as the faulty statement, ran ssFix's code search twice (once for each statement),
and used the better result.}. The code database we used consists of the 
DARPA MUSE code repository (as used in \Cref{test_fi_expt}) as the external 
code repository and the same five projects used in \cite{xin17leveraging} as 
the local programs. We filtered
away candidates that are syntactically redundant and those that are simply from
the bug-fixed versions. We looked at the top-500 candidates as the retrieval 
results.

\subsubsection{Result} \label{eval_ssfix_codesearch_rslt}

\begin{figure}
\centering
\includegraphics[width=0.313\textwidth,height=0.2\textwidth]{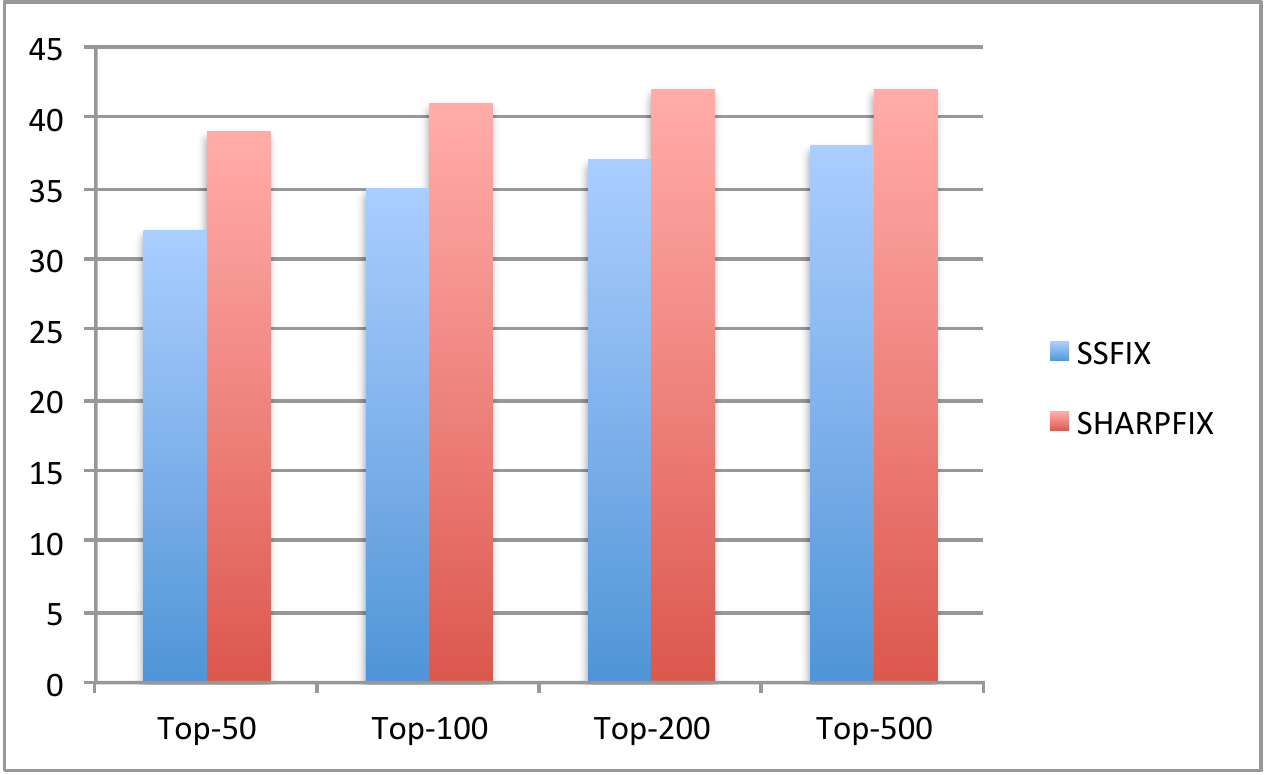}
\caption{The Retrieval of Candidates that Contain the Fix Ingredients
(We looked at the top-50, 100, 200, and 500 candidates. The column shows the
number of bugs for which ssFix/sharpFix retrieved promising candidates)}
\label{cs_rslt0}
\vspace{-1.5em}
\end{figure}

\Cref{cs_rslt0} shows the numbers of promising candidates ssFix retrieved
within the top-$k$ results (with $k$ being 50, 100, 200, and 500 respectively).
Within the top-500 results, ssFix retrieved promising candidates for 38 bugs:
it retrieved in total 61 candidates that contain the fix ingredients in the 
parameterized forms, among which, 38 candidates are promising, i.e., contain
the exact fix ingredients after translation. In \Cref{test_fi_expt_rslt},
we showed that for as many as 80 bugs, the fix ingredients in the parameterized
forms exist. ssFix retrieved promising fix ingredients for 38/80=47.5\% bugs.

\subsection{Evaluating ssFix's Code Reuse} \label{eval_ssfix_codereuse}

To evaluate ssFix's code reuse ability, we looked at the 61 bugs for which ssFix
retrieved candidates that contain the fix ingredients in the parameterized forms.
For each bug, we provided ssFix with the target it produced and the candidate
it retrieved, and ran ssFix's patch generation and patch validation automatically.
If ssFix produced a plausible patch, we manually checked whether the patch is correct 
(i.e., semantically equivalent to the developer patch provided by Defects4J).

Our results show that ssFix produced 25 plausible patches among which 23 are
correct. It successfully reused 23/61=37.7\% candidates for bug repair. Note that
this is a lower bound because not all the candidates can be reasonably reused 
for repair. We found that the exact fix ingredients (without any
translation) are contained in 38 candidates, and we expect ssFix to be able to
reuse those fix ingredients in producing the correct patches. For the other 23 
(61-38) candidates which only contain the fix ingredients in the parameterized forms, 
we manually determined whether they can be reasonably reused for bug repair.
We identified only 3 such candidates (it may not be reasonable for a repair
technique to translate an arbitrary, parameterized fix ingredient
into the exact one to be reused for repair). 

We analyzed the failures of ssFix in reusing the 18 (38+3-23) reasonable 
candidates for producing the correct patches.
We found that 7 candidates are not ideal for repair.
As one example, for the bug Cl119, ssFix retrieved a candidate containing
the exact fix ingredient \texttt{case Token.CATCH:} as a statement to be inserted 
in the target for producing the correct patch. However, the fix ingredient is embedded
in a big switch statement, and it is therefore difficult for ssFix to leverage 
the fix ingredient to do the correct repair.
For the other failures, we found that ssFix yielded bad candidate
translations for 3 cases, it created bad code matching results
for 2 cases, and its modifications are not sophisticated enough for producing
the correct patches for 6 cases. We identified the key shortcomings of
ssFix in translation, code matching, and modification, and developed sharpFix
for an improvement. More details can be found in \Cref{sharpfix_codereuse}.

\subsection{Evaluating ssFix's Repair} \label{eval_ssfix_repair}

We conducted three experiments to evaluate ssFix's repair abilities:
one experiment (E0) to evaluate ssFix's full repair ability and two more 
experiments (E1 \& E2) to evaluate its partial repair abilities with 
the fault-located statement and method manually provided.

\subsubsection{Experiment}

For E0, we ran ssFix to repair all the 357 Defects4J bugs automatically.
For E1, we manually identified 112 Defects4J bugs\footnote{Note that the
bugs used for E1 are not the same ones used in \Cref{test_fi_expt}. Here we 
looked for bugs for which we can identify single faulty statements. The
developer patches for these bugs are not necessarily simple.} for repair. For each bug,
the developer patch makes changes for a single statement, we manually identified 
the statement\footnote{We identified the statement as the first enclosing 
statement where a fixing change was made. If there are multiple such statements, 
we ignored the bug.}, and provided ssFix with it as the faulty statement for
bug repair. In the case where the developer patch inserts a statement, 
we identified its two adjacent statements (at most) in the inserted statement's block,
considered each adjacent statement as the faulty statement, ran ssFix to repair
each, and used the better result. For E2, we manually identified 201 bugs.
For each bug, the developer patch makes changes within only one method.
We manually identified the method, ran ssFix's fault localization to obtain a 
list of suspicious statements within it, and provided the list of statements 
to ssFix for repair. ssFix used the same code database used in
\Cref{eval_ssfix_codesearch_expt} for
bug repair. It ignored any candidates that are from the bug-fixed versions of 
the faulty programs. We set the maximum number of candidates used by ssFix for 
repairing each suspicious statement to be 200. We set the time budget and memory 
budget for repairing each bug as two hours and 8GB for all experiments. We ran 
all the experiments on a machine with 32 Intel-Xeon-2.6GHz CPUs and 128GB memory.

\subsubsection{Results}

\begin{table}[t]
\centering \tiny
\caption{The Results of E0 (the Full Repair Experiment)}
\label{e0_rslt}
\resizebox{.48\textwidth}{!}{%
\begin{threeparttable}
\begin{tabular}{|c|c|c|c|c|c|c|c|c|c|c|c|c|} \hline
\multirow{3}{*}{\textbf{\parbox{0.5cm}{Project\\(\#Bugs)}}} & \multicolumn{6}{c|}{\textbf{sharpFix}} &
\multicolumn{6}{c|}{\textbf{ssFix}} \\ \cline{2-13} & \multicolumn{4}{c|}{\textbf{Time
(min.)}} & \multirow{2}{*}{\textbf{\#P}} & \multirow{2}{*}{\textbf{\#C}} & \multicolumn{4}{c|}{\textbf{Time (min.)}} & \multirow{2}{*}{\textbf{\#P}} & \multirow{2}{*}{\textbf{\#C}} \\\cline{2-5} \cline{8-11} & \textbf{Min} & \textbf{Max} & \textbf{Med} & \textbf{Avg} & & & \textbf{Min} & \textbf{Max} & \textbf{Med} & \textbf{Avg} & & \\ \hline
C (26) & 0.8 & 115.7 & 7.2 & 19.2 & 9 & 4 & 1 & 80.7 & 12.4 & 20.7 & 7 & 2 
\\ \hline
Cl (133) & 1.8 & 96.1 & 21.3 & 26 & 17 & 4 & 2.5 & 54.9 & 10.1 & 16.3 & 14 & 2 
\\ \hline
M (106) & 0.7 & 118.5 & 11.3 & 33.2 & 33 & 13 & 1 & 119.3 & 14.7 & 30.2 & 26 & 8 
\\ \hline
T (27) & 1.6 & 30 & 12.2 & 15.1 & 5 & 0 & 1.4 & 37.3 & 7.5 & 13.5 & 4 & 0 
\\ \hline
L (65) & 0.8 & 116.1 & 4.8 & 18 & 25 & 15 & 0.8 & 117.8 & 4.3 & 13.1 & 18 & 10 
\\ \hline \hline
Sum (357) & 0.7 & 118.5 & 11.3 & 25.1 & 89 & 36 & 0.8 & 119.3 & 10.1 & 21 & 69 & 22
\\ \hline
\end{tabular}
\begin{tablenotes}
\item We show the projects in their abbreviations: C is JFreeChart; Cl is Closure
Compiler; M is Commons Math; T is Joda-Time; and L is Commons Lang. \#P and \#C are 
the respective numbers of the plausible and correct patches generated.
\end{tablenotes}
\end{threeparttable}}
\vspace{-3em}
\end{table}

The results of E0 are shown in \Cref{e0_rslt} (see the right six columns).
ssFix produced in total 69 plausible patches with the median and average times 
of producing a plausible patch being about 
10 and 21 minutes respectively. Among the 69 patches, 22 are correct. 
We manually determined the correctness of a plausible patch by comparing it to 
the developer patch in the dataset and checking whether the two patches are 
semantics-equivalent. We explained for each correct patch why it is 
semantics-equivalent to the developer patch (see the link we provided 
at the end of \Cref{eval_rslts}).

Our results of E1 show that ssFix produced 42 plausible patches among which
26 are correct. With the faulty statement known in advance, the median
and average times for producing a plausible patch are 1.9 and 3 minutes. 
For the same 112 bugs in E0, the fully automatic repair experiment, ssFix 
produced plausible and correct patches for 38 and 21 bugs. With the faulty 
statement being accurately located, ssFix correctly repaired (26-21)/21=23.8\% 
more bugs.

Our results of E2 show that ssFix produced 61 plausible patches among which
27 are correct. With the faulty method known in advance, the median and average 
times for producing a plausible patch are 3.5 and 13 minutes. For the
same 201 bugs in E0, ssFix produced plausible
and correct patches for 54 and 22 bugs. With the faulty method being accurately
located, ssFix correctly repaired (27-22)/22=22.7\% more bugs. Our results of
E1 \& E2 show that more accurate fault localization can significantly improve a
technique's repair ability.

\section{sharpFix} \label{sharpfix}

We identified possible weaknesses of ssFix's approaches, and developed
the new APR technique sharpFix\footnote{sharpFix was built upon ssFix's
implementation available at \textit{https://github.com/qixin5/ssFix}.} 
as an improvement. We conducted similar experiments to evaluate sharpFix. 
The results demonstrated that compared to ssFix, sharpFix has better code 
search, code reuse, and repair abilities. In this section, we elaborate 
on how sharpFix works and show the experimental results we got.

\subsection{Overview}

Similar to ssFix, sharpFix is an APR technique that reuses existing code fragments 
from a code database for bug repair. It also goes through four stages:
\textit{fault localization}, \textit{code search}, \textit{patch generation}, and
\textit{patch validation} to possibly produce a plausible patch. sharpFix uses
ssFix's method to do fault localization. Its methods for code search and code reuse
(i.e., patch generation and patch validation) however are different from ssFix's. 
For code search, sharpFix uses different methods for searching candidates 
from the local faulty program and from the code repository and combines the results
(i.e., the retrieved candidates) for reuse. For patch generation, sharpFix
goes through the same steps used by ssFix: \textit{code translation}, 
\textit{code matching}, and \textit{modification}, each step however is different.
For patch validation, sharpFix's method is identical to ssFix's method except that
sharpFix performs S6's static resolving technique \cite{reiss09semantics} as an extra
step to check the validity of a patch (e.g., whether it used an undeclared variable) 
before dynamically validating the patched program.

We use the Defects4J bug C4 (\Cref{c4_eg}) as an example to explain how 
sharpFix works. For this bug,
the developer patch produced an if-condition (Line 1) to guard the three 
statements (Lines 2-4) to avoid a null-pointer
exception. Though the repair is relatively easy, ssFix and many other existing 
techniques \cite{legoues12,z_qi15,xuan16,le16,xiong17,chen17contract} failed 
to produce the correct patch. sharpFix successfully produced the correct patch 
in about 9.7 minutes\footnote{
sharpFix is not specifically designed to repair null-pointer errors. A repair 
technique that targets on such errors (e.g., \cite{cornu15npefix}) may work faster.}. 
For this bug, sharpFix's (also ssFix's) fault localization identified the 
statement at Line 2 as top-suspicious statement for repair.

\begin{figure}
\centering
\begin{lstlisting}
+ if (r!=null) {
    Collection c = r.getAnnotations();                        
    Iterator i = c.iterator();
    while (i.hasNext()) { ... }
+ }
\end{lstlisting}
\caption{The Developer Patch for the C4 Bug}
\label{c4_eg}
\vspace{-.5cm}
\end{figure}

\subsection{Code Search} \label{sharpfix_cs}

\subsubsection{Improving ssFix's Code Search} \label{sharpfix_cs_motiv}

For code search, ssFix uses relatively small code chunks (including at 
most three statements) and uses the same searching method for both local 
(within the local faulty program) code search and global (within the 
code repository) code search. It is possible to obtain better code search 
results using code chunks that are smaller or larger (than ssFix's code 
chunks) and using different searching methods for local and global code 
search (a better way to do local code search, compared to global code search, 
is to use less parameterization). To investigate these possibilities, we
conducted experiments comparing a set of different code search methods.

More specifically, we re-ran ssFix's the code search experiments
in \Cref{eval_ssfix_codesearch_expt} using different sizes of code chunks:
we created six code search methods K5WC1, K5WC2, K5WC3, K5WC4, K5WC5, and
K5WMD that look at using code chunks containing one to five statements 
(K5WC1 to K5WC5), and the whole Java method (K5WMD). These search methods 
are equivalent to ssFix's search method except that the used chunks are in 
different sizes. For each search method, we counted the number of bugs whose 
exact fix ingredients can be found from any retrieved candidate (with and 
without ssFix's translation\footnote{We additionally checked whether the 
exact fix ingredient can be found in the original untranslated candidate 
because it is possible for ssFix to yield a bad translation.}) in the top-50, 
100, 200, and 500 results. We found that K5WMD overall yielded the best result: 
It retrieved 44 candidates that contain the fix ingredients from the top-500 
results.

ssFix extracts structural tokens as $k$-grams (where $k$=5) from the target and
the candidate for calculating their similarity score. We investigated using 
different values of $k$ for the best search method K5WMD: we created two more
search methods K3WMD and K7WMD and ran ssFix's code search experiment with each. 
We found that K3WMD yielded the best results. It retrieved the most candidates 
containing fix ingredients, though it is not significantly better than the other 
two methods. Compared to the second-best method K5WMD, K3WMD retrieved the same 
amount of candidates that contain the fix ingredients within the top-50 and 
top-500 results, and it retrieved only one more candidate within the top-100 
and top-200 results each.

We investigated three search methods: LCS0, LCS1, and LCS2 that we developed for
local code search. Given a code chunk (either the target or the candidate), the
methods extract different tokens as shown below.
\begin{itemize}
\item \textbf{LCS0}: The tokens are the textual contents (in the compacted
forms with no whitespaces) of all the children expressions of the chunk statement(s).
\item \textbf{LCS1}: The tokens are all the conceptual tokens that ssFix extracts
(using its own method in Section \Romannum{3}-A(2) of \cite{xin17leveraging}) from the
chunk statement(s) (with the Java keywords, stop words, and short and long words 
filtered by ssFix kept).
\item \textbf{LCS2}: The tokens are all the conceptual tokens that ssFix extracts
from the chunk statement(s) (with the Java keywords, stop words, and short and long 
words kept) plus any symbols that are non-Java-identifiers (e.g., \textit{+=}).
\end{itemize}
Given the tokens (as two lists) extracted from the target and the candidate, 
the search method does token matching and uses the Dice Similarity\footnote{We 
slightly changed the original measure to be used for lists.} to
calculate a similarity score.
For each search method, we show below as an example, the extracted tokens
(in angle brackets) for the statement at Line 2 in \Cref{c4_eg}.
\begin{lstlisting}[numbers=none]
LCS0 (7 in total):
<Collectionc=r.getAnnotations();>, <Collection>, <c=r.getAnnotations()>, 
<c>, <r.getAnnotations()>, <r>, <getAnnotations()>

LCS1 (7 in total):
<collection>, <collect>, <c>, <r>,
<getannotations>, <get>, <annotations>

LCS2 (12 in total):
<collection>, <collect>, <c>, <=>, <r>, <.>, 
<getannotations>, <get>, <annotations>, <(>, <)>, <;>
\end{lstlisting}
Note that the code chunk (either the target or the candidate) we used for local
code search contains only one statement. This is because a code chunk containing 
multiple, sequential statements is more likely to be unique within the local 
program (than it is within a large code repository). To evaluate the three local
search methods, we ran each method to
retrieve candidates within each local faulty program of the 103 bugs (with simple 
patches) and checked whether the fix ingredient (with and without ssFix's 
translation) can be
found in any candidate from the top results. Our results showed that LCS1 yielded 
the best result: it retrieved 22 candidates containing the fix ingredients from 
the top-500 results, though not significantly better than LCS0 (the second best
method). By looking at the fix ingredients, we found that the best local search
method LCS1 can complement the best global search method K3WMD in finding 5 (11.4\%) 
more fix-ingredient-contained candidates.

\subsubsection{Methodology}

In \Cref{sharpfix_cs_motiv}, we showed that K3WMD and LCS1 yielded the best
searching results for global and local code search respectively and that it
is possible to combine the two to yield better results. Based on these findings,
we developed sharpFix's code search method which performs K3WMD and LCS1
to do code search separately and merges the results together to obtain a list 
of candidates to be reused for bug repair.

More specifically, given a suspicious statement $s$, sharpFix produces two
code chunks: $tchunk_0$ which contains the enclosing method of $s$ and 
$tchunk_1$ which contains $s$ itself. Next it performs K3WMD using $tchunk_0$
as the query to obtain a list of ranked candidate code chunks $cchunks_0$
from the code repository and it performs LCS1 using $tchunk_1$ as the query to 
obtain a list of ranked candidate code chunks $cchunks_1$ from the local faulty 
program. For each $cchunk_0\in cchunks_0$ which is actually a Java method, 
sharpFix does not simply use the whole method for code reuse but identifies 
small code fragments as single statements within the method that are likely to
be relevant to $s$ (and are thus useful for repair). To obtain such code fragments, 
sharpFix first translates $cchunk_0$ into $rcchunk_0$
(using the translation method we will explain in \Cref{sharpfix_code_translation}),
and uses LCS1 to identify two statements $cs_0$ and $cs_1$ in $rcchunk_0$ that
are most similar to $s$. Each of the two statements is associated with the
searching score of $cchunk_0$. Next sharpFix normalizes separately the searching 
scores of (1) the statement candidates identified from $cchunks_0$ and (2) the 
candidates from $cchunks_1$, merges the candidates together, and ranks them by 
the normalized scores. This way, sharpFix obtains a list of candidates 
(as statements) retrieved from both the local faulty program and the code
repository. For code reuse, sharpFix looks at each candidate in the list to 
possibly produce a plausible patch for the target which contains the suspicious
statement $s$ only.

\begin{figure}
\centering
\begin{lstlisting}
if (r!=null) {
  result = r.getUpperBound();
}
\end{lstlisting}
\caption{A Candidate for the C4 Bug}
\label{c4_cand}
\vspace{-.5cm}
\end{figure}

For our example shown in \Cref{c4_eg}, given the suspicious statement (at Line 2)
identified by fault localization, sharpFix performed code search and retrieved a
candidate (rank 105) from the local faulty program that contains the single 
statement shown at Line 2 in \Cref{c4_cand}.

\subsubsection{Evaluation} \label{eval_sharpfix_codesearch_expt}

We conducted an experiment similar to the one used in \Cref{eval_ssfix_codesearch_expt}
to evaluate sharpFix.
For each of the used 103 Defects4J bugs, we provided sharpFix with the faulty
statement, ran its code search, and checked whether the fix ingredient was contained
in the retrieved candidate statements. As we did in \Cref{eval_ssfix_codesearch_expt}, 
we filtered away candidates that are syntactically redundant\footnote{We looked at 
the candidate statement plus its two neighbouring statements (which sharpFix uses
for insertion) to identify redundancy.} and those that are simply from the bug-fixed 
versions, and we
looked at the top-500 retrieved candidates. For each candidate (that contains a 
single statement), we performed sharpFix's translation to translate its enclosing 
method, and checked whether the exact fix ingredient is contained in the translated 
statement, its two neighbouring statements (which sharpFix uses for insertion), and 
the enclosing if-condition when the enclosing statement is an if-statement (sharpFix 
uses the enclosing if-condition to produce a new if-statement for patch generation).

\Cref{cs_rslt0} shows our results (see the red bars).
We found that sharpFix's code search method is better than ssFix's. 
It retrieved promising candidates that contain the exact
fix ingredients for 39 bugs within the top-50 results and for 42 bugs within the
top-500 results. Compared to ssFix's code search, though sharpFix's code search 
retrieved only four more promising candidates within the top-500 results, it retrieved 
39 promising candidates within the top-50 results which are more than all the
promising candidates ssFix retrieved within the top-500 results. 
\Cref{test_fi_expt_rslt} showed that the parameterized fix ingredients exist for
80 bugs. sharpFix retrieved promising fix ingredients for 42/80=52.5\% bugs.

\subsection{Code Reuse} \label{sharpfix_codereuse}

We analyzed ssFix's failures in its code reusing steps: code translation,
code matching, and modification in \Cref{eval_ssfix_codereuse}, identified
possible ways for improvement, and developed sharpFix's method for each step.

\subsubsection{Code Translation} \label{sharpfix_code_translation}

For code translation, ssFix first maps identifiers in the candidate (or candidate
identifiers) to the related identifiers in the target (or target identifiers), and 
then renames the candidate identifiers as the mapped target identifiers to yield a
translated candidate. We identified two weaknesses of ssFix's current method
for mapping identifiers: (1) ssFix only maps candidate identifiers to identifiers
that are used in the target. A candidate identifier can be related to an identifier
that is not used in the target but is accessible there; and (2) ssFix identifies 
two identifiers as related only based on matching the code patterns of their
usage contexts. This can be insufficient: Two identifiers can be highly related but
their usage contexts are not identical (though very similar). Identical usage 
contexts may also not imply that the two identifiers being compared are the most 
related.

\begin{algorithm}
\tiny
\caption{Creating an Identifier Mapping} \label{imap-alg}
\begin{algorithmic}[1]
\Require $tchunk$, $cchunk$
\Ensure $imap$[$id\rightarrow id$] \Comment{$id$ is an identifier binding}
\State $imap$[$id\rightarrow id$]$\gets$ empty
\State $cids\gets$ collect all the non-JDK candidate identifiers (those appear in $cchunk$'s method)
\State $tids\gets$ collect all the non-JDK target identifiers (those appear in $tchunk$'s method and appear as the declared fields and methods of $tchunk$'s class)
\ForAll{$cid\in cids$}
  \If{!shortName($cid$)} \Comment{the string length is greater than 2}
    \State $tid\gets$ find the first $tid\in tids$ whose name is equal to $cid$'s name
    \If{$tid$ exists and is compatible with $cid$}
      \State $imap$.add($cid$,$tid$)
    \EndIf
  \EndIf
\EndFor
\State $cmid\gets$ get $cchunk$'s enclosing method identifier
\State $tmid\gets$ get $tchunk$'s enclosing method identifier
\State $ccid\gets$ get $cchunk$'s enclosing class identifier
\State $tcid\gets$ get $tchunk$'s enclosing class identifier
\If{!$imap$.containsKey($ccid$) \&\& !$imap$.containsValue($tcid$)}
  \State $imap$.add($ccid$,$tcid$)
\EndIf
\If{!isConstructorId($cmid$) \&\& !$imap$.containsKey($cmid$) \&\& !$imap$.containsValue($tmid$)}
  \State $imap$.add($cmid$,$tmid$)
\EndIf
\State $imap\gets$ mapIdsByContexts($cids$,$tids$,$imap$) \Comment{using ssFix's method}
\State $umcids\gets$ get all $cid$s that are currently unmapped
\State $umtids\gets$ get all $tid$s that are currently unmapped
\ForAll{$umcid\in umcids$}
  \State $bestmatch\gets$ find the $umtid$ in $umtids$ that is compatible with $umcid$ and share the most conceptual tokens with $umcid$ (measured by Dice Similarity)
  \If{$bestmatch$ exists}
    \State $imap$.add($umcid$,$bestmatch$)
  \EndIf
\EndFor
\State \Return $imap$
\end{algorithmic}
\end{algorithm}

To address the two problems, we developed a new algorithm (\Cref{imap-alg})
used by sharpFix to create an identifier mapping. Based on the created
identifier mapping, sharpFix renames each candidate identifier as its mapped 
target identifier to produce a translated candidate. To address (1), sharpFix
looks for candidate identifiers that appear in the candidate's enclosing method
and it looks for target identifiers that not only appear in the target but also 
(a) appear in the target's enclosing method, (b) appear as the declared fields of
the target's class, and (c) appear as the declared methods of the target's class.
sharpFix looks for such candidate identifiers (that appear in the candidate's 
enclosing method) because it may actually use those identifiers to produce a 
patch (which we will explain in \Cref{sharpfix_mod}). sharpFix looks for target 
identifiers from (a), (b), and (c) to produce a set of accessible identifiers in 
the target to which a candidate identifier can be mapped. To address (2), sharpFix 
identifies related identifiers based on not just their usage contexts but their 
string lengths, string equality, locations, and shared concepts (measured by the 
overlap of the extracted conceptual tokens).

As shown in \Cref{imap-alg}, sharpFix accepts the target chunk $tchunk$ and
the candidate chunk $cchunk$ as input. As output, sharpFix creates an identifier
mapping $imap$. sharpFix starts by collecting two lists of non-JDK identifiers
$cids$ and $tids$ (Lines 2 \& 3) which are actually identifier bindings
(e.g., representing a variable declaration and its use). It first maps
identifiers that share the same names which are not too short (Lines 4-8),
next maps the method and class identifiers (Lines 9-16), next maps identifiers
by their usage contexts (Line 17), and finally maps identifiers by the
shared conceptual words extracted in their names (Lines 18-23). sharpFix
uses ssFix's method (from Section \Romannum{3}-A(2) of \cite{xin17leveraging})
for matching two identifiers' usage contexts and for extracting conceptual words.

The translated version yielded by sharpFix for the candidate shown in \Cref{c4_cand}
is just as itself (sharpFix changed the candidate's enclosing method's name and
the name of a method call that do not appear in the candidate and are not shown).
For the variable identifier \texttt{result} in the candidate, sharpFix found an
identifier in the
target's method that has the same name, mapped \texttt{result} to this identifier, 
and renamed \texttt{result} as itself. For the variable \texttt{r} in the candidate, 
sharpFix mapped it to the identifier \texttt{r} in the target based on their matched
usage contexts (e.g., as both \texttt{r!=null}). sharpFix did not map 
\texttt{getUpperBound} to any target identifier and thus did not change it 
(it did not map \texttt{getUpperBound} to \texttt{getAnnotations} since they only 
share a stop word \texttt{get}).

\subsubsection{Code Matching}

ssFix's code matching method is based on matching rules and arbitrary thresholds. 
We found this makes ssFix's code matching somewhat inflexible.
For example, it does not allow two method calls to match unless the method
names are identical which can sometimes hinder ssFix from repairing an
incorrectly called method. sharpFix uses a new method to do code matching which
uses simplified matching rules and no thresholds. It matches statements/expressions 
based on the extracted conceptual tokens and symbols, i.e., the LCS2 tokens shown 
in \Cref{sharpfix_cs_motiv}. To do code matching, sharpFix accepts the target 
$tchunk$ and the translated candidate $rcchunk$ as input.
As output, it produces a code mapping $cmap$ that maps each statement/expression 
in $tchunk$ to its matched statement/expression in $rcchunk$.
To create such a mapping, sharpFix starts by collecting two lists of statements and 
expressions $tses$ and $cses$ from $tchunk$ and $rcchunk$ respectively (by visiting
the ASTs in pre-order). The collected expressions are non-trivial and do not 
include identifiers, number constants, or any of the four types of literals: 
\textit{boolean}, \textit{null}, \textit{character}, and \textit{string}.
For each statement/expression $tse$ in $tses$, sharpFix finds a $cse$ in $cses$ that 
is compatible with $tse$ and shares the most LCS2 tokens with $tse$ (measured by
the Dice Similarity) and maps $tse$ to $cse$.

When two $se$s (statements/expressions) are both statements, they are compatible
if they are both loops. Otherwise, they need to have the same statement 
type\footnote{The type of a statement/expression is its corresponding node type 
in the abstract syntax tree that sharpFix builds using the Eclipse JDT library 
\cite{eclipsejdt}.}
(e.g., both as \textit{return} statements) to be compatible.
When two $se$s are both expressions, they are compatible if their expression types 
are equal. When one $se$ is a statement and the other is an expression, they are 
only compatible if the statement's type is \textit{VariableDeclarationStatement} and 
the expression's type is either \textit{Assignment} or 
\textit{VariableDeclarationExpression}.

For the bug example, sharpFix maps the target statement at Line 2 in \Cref{c4_eg}
to the matched (also translated) candidate statement at Line 2 in \Cref{c4_cand}. 
The extracted tokens and the similarity calculation are 
shown below.
\begin{lstlisting}[numbers=none]
LCS2 Tokens from the Target Statement (12 in total):
<collection>, <collect>, <c>, <=>, <r>, <.>, 
<getannotations>, <get>, <annotations>, <(>, <)>, <;>

LCS2 Tokens from the Candidate Statement (11 in total):
<result>, <=>, <r>, <.>, 
<getupperbound>, <get>, <upper>, <bound>, <(>, <)>, <;>

Overlapped Tokens (7 in total):
<=>, <r>, <.>, <get>, <(>, <)>, <;>

Dice Similarity: (2*7)/(12+11)=0.609
\end{lstlisting}

\subsubsection{Modification} \label{sharpfix_mod}

ssFix uses three types of modifications: replacement, insertion, and deletion
to produce patches based on the matched and unmatched statements/expressions
between the target and the translated candidate. We made sharpFix's modification 
strategy more sophisticated by adding two more modifications: 
\textit{adding if-guard} and \textit{method replacement}. To produce patches using
\textit{adding if-guard}, sharpFix looks at a target statement $s$ (which
appears in the target) and its mapped candidate statement $s'$ (which appears
in the translated candidate). If the parent of $s'$ is an if-statement with a 
condition $e'$, sharpFix creates new if-statements with the condition $e'$
to guard existing statements in the target (and possibly in its enclosing method).
Currently, sharpFix selects two sets of statements to be guarded: (1) the 
target statement $s$ itself and (2) $s$ plus the following statements its block, 
and produces the corresponding patches. To produce a patch using the
modification \textit{method replacement}, sharpFix replaces the enclosing method
of the target with the enclosing method of the translated candidate to possibly 
support making multiple changes in the scope of a method.

sharpFix uses the same method used by ssFix to do replacement. For insertion,
sharpFix looks at the candidate statement $s'$ (translated) to which the target 
statement $s$ is mapped, identifies the adjacent statements of $s'$ in its block:
$s_0'$ and $s_1'$ (translated) that are before and after $s'$, and inserts 
$s_0'$ before $s$ and $s_1'$ after $s$ to yield two patches. sharpFix does not use 
ssFix's method to do insertion
because the target and candidate chunks it looks at both contain only one statement.
sharpFix does not use ssFix's deletion because it was shown in \cite{xin17leveraging} 
to be likely to produce defective patches.

For the bug example, sharpFix produces two patches using the modification
\textit{adding if-guard}. As one patch, sharpFix uses the if-condition
\texttt{r!=null} to guard the target statement (Line 2 in \Cref{c4_eg}) only.
The patched program however fails to compile because variable \texttt{c} at 
Line 3 becomes undeclared. As the other patch, sharpFix uses the if-condition 
to guard the target statement plus its following statements in the block.
This is the correct patch shown in \Cref{c4_eg}.

\subsubsection{Evaluation}

sharpFix retrieved candidates that contain the parameterized fix ingredients 
for 59 bugs within the top-500 results. To evaluate sharpFix's code reuse, we 
looked at the 59 bugs. For each bug, we provided sharpFix with the target and 
the retrieved
candidate, and ran sharpFix's patch generation and patch validation automatically. 
If sharpFix produced a plausible patch, we manually checked whether the patch is 
correct. Our results show that sharpFix produced 30 plausible patches which are all 
correct, and successfully reused 30/59=50.8\% candidates for bug repair.

The exact fix ingredients (without any translation) are contained in 39 candidates,
and we expect sharpFix to be able to reuse those fix ingredients in producing the 
correct patches. For the other 20 (59-39) candidates which only contain the fix 
ingredients in the parameterized forms, we identified only three candidates that
can be reasonably reused for repair. We analyzed sharpFix's failures in reusing
the candidates for repairing the 12 (39+3-30) bugs and found that the candidates
are not ideal for repairing 9 bugs (an example of such candidate can be found
in \Cref{eval_ssfix_codereuse}). To successfully reuse the candidates to repair
the other 3 bugs, sharpFix's modification needs to be more sophisticated. Though 
we can make sharpFix's modification more sophisticated, doing so may not actually 
improve its overall repair performance (as suggested in \cite{long16-anal}).

\subsection{Repair} \label{sharpfix_repair}

We also conducted the three experiments (E0, E1, and E2) used in 
\Cref{eval_ssfix_repair} to evaluate sharpFix's repair abilities. 
\Cref{e0_rslt} shows the results of E0: the full repair experiment.
sharpFix produced in total 89 plausible patches (for 89 bugs) among 
which 36 are correct. For E1 and E2, with the faulty statement and
method known in advance, sharpFix produced correct patches for
34.5\% and 19.4\% more bugs respectively (compared to the results
of E0).

\section{Evaluation}

We evaluated sharpFix, ssFix, and four other repair techniques
jGenProg \cite{astorimpl}, jKali \cite{astorimpl}, Nopol (version
2015) \cite{nopolimpl}, and HDRepair \cite{hdrepairimpl} on a
different bug dataset: Bugs.jar-ELIXIR \cite{saha17elixir} which
is a sample of the Bugs.jar dataset \cite{saha18bdj} and contains
127 real bugs. Our results confirm that sharpFix is an improvement
over ssFix and show that it outperformed the others in successfully 
repairing more bugs.

\subsection{Setup}

The Bugs.jar dataset \cite{saha18bdj} consists of 1,158 real Java bugs
drawn from 8 open-source Java projects. The Bugs.jar-ELIXIR dataset \cite{saha17elixir}
created by Saha et al. is a sample of the original dataset and contains 
127 real bugs drawn from 7 of the 8 Java projects. For each of the 127 bugs, 
the bug-fixing change is local (within a hunk). Although these bugs are 
relatively easy for repair (since their bug-fixing changes are local and are 
thus simple), we think they are still reasonable to be used for evaluating 
existing APR techniques since no existing APR technique has been shown to 
be good at repairing complex bugs through making complex, non-local fixes.

For evaluation, we ran sharpFix, ssFix, and all other techniques each to repair 
all the 127 bugs. The external code repository used by sharpFix and ssFix is
the DARPA MUSE repository. ssFix uses the earliest versions of the 7 projects
as the local projects. The time and memory budgets used by each technique for
repairing a bug are two hours and 8GB. All the experiments were run on the same
machine we used for the experiments shown in \Cref{eval_ssfix_repair,sharpfix_repair}.
Given that jGenProg and HDRepair use randomness for patch generation, we ran
each technique in three trials\footnote{Running jGenProg and HDRepair only in 
three trials might be insufficient to show the tools' full abilities. 
However, we believe our results are sufficient to show that sharpFix outperforms 
these tools: it generated more than 10 correct patches in one trial than these 
tools did in three trials.} to repair a bug.
We did not compare sharpFix to many other repair techniques that are written for C
(e.g., SearchRepair \cite{ke15}, CodePhage \cite{douskos15}, Prophet \cite{long16}, 
and Angelix \cite{mechtaev16}) or are not publicly available (e.g., PAR \cite{kim13}) 
including ELIXIR \cite{saha17elixir}.

\subsection{Results} \label{eval_rslts}

\begin{table}[t]
\centering \tiny
\caption{Comparing sharpFix \& ssFix on Repairing Bugs.jar Bugs} 
\label{bugs_dot_jar_eval_table0}
\resizebox{.48\textwidth}{!}{%
\begin{threeparttable}
\begin{tabular}{|c|c|c|c|c|c|c|c|c|c|c|c|c|} \hline
\multirow{3}{*}{\textbf{\parbox{0.5cm}{Project\\(\#Bugs)}}} & \multicolumn{6}{c|}{\textbf{sharpFix}} &
\multicolumn{6}{c|}{\textbf{ssFix}} \\ \cline{2-13} & \multicolumn{4}{c|}{\textbf{Time
(min.)}} & \multirow{2}{*}{\textbf{\#P}} & \multirow{2}{*}{\textbf{\#C}} & \multicolumn{4}{c|}{\textbf{Time (min.)}} & \multirow{2}{*}{\textbf{\#P}} & \multirow{2}{*}{\textbf{\#C}} \\\cline{2-5} \cline{8-11} & \textbf{Min} & \textbf{Max} & \textbf{Med} & \textbf{Avg} & & & \textbf{Min} & \textbf{Max} & \textbf{Med} & \textbf{Avg} & & \\ \hline
ACC (10) & 1.2 & 1.2 & 1.2 & 1.2 & 1 & 1 & 1.3 & 4.1 & 2.7 & 2.7 & 2 & 1
\\ \hline
CML (16) & 39.2 & 46.5 & 42.9 & 42.9 & 2 & 1 & 35.6 & 118.1 & 69.6 & 73.3 & 4 & 2
\\ \hline
FLK (7) & 0.8 & 0.8 & 0.8 & 0.8 & 1 & 1 & 6.9 & 6.9 & 6.9 & 6.9 & 1 & 1
\\ \hline
OAK (31) & 2.6 & 98.5 & 11.2 & 28.3 & 10 & 0 & 0.6 & 111.2 & 5.4 & 21.4 & 14 & 1
\\ \hline
MAT (21) & 0.8 & 103.4 & 26.2 & 32.2 & 10 & 6 & 0.8 & 64.9 & 11.5 & 17.2 & 9 & 5
\\ \hline
MNG (5) & 32.4 & 32.4 & 32.4 & 32.4 & 1 & 0 & 0.6 & 0.6 & 0.6 & 0.6 & 1 & 0
\\ \hline
WCT (37) & 0.9 & 91.4 & 8.4 & 21.6 & 14 & 6 & 3 & 82.8 & 8.7 & 22.1 & 12 & 1
\\ \hline \hline
Sum (127) & 0.8 & 103.4 & 12.8 & 26.3 & 39 & 15 & 0.6 & 118.1 & 8.5 & 23.9 & 43 & 11
\\ \hline
\end{tabular}
\begin{tablenotes}
\item We show the projects in their abbreviations: ACC is Accumulo; CML is Camel; 
FLK is Flink; OAK is Jackrabbit Oak; MAT is Commons Math; MNG is Maven; and WCT is 
Wicket. \#P and \#C are the respective numbers of the plausible and correct patches 
generated.
\end{tablenotes}
\end{threeparttable}}
\vspace{-1.5em}
\end{table}

\begin{table}[t]
\centering \tiny
\caption{Comparing All Techniques on Repairing Bugs.jar Bugs} 
\label{bugs_dot_jar_eval_table1}
\begin{threeparttable}
\begin{tabular}{|c|c|c|c|c|c|c|} \hline
\multirow{2}{*}{\textbf{Tool}} & \multicolumn{4}{c|}{\textbf{Time (min.)}} & \multirow{2}{*}{\textbf{\#Plausible}} & \multirow{2}{*}{\textbf{\#Correct}} \\ \cline{2-5}
& \textbf{Min} & \textbf{Max} & \textbf{Med} & \textbf{Avg} & & \\ \hline
sharpFix & 0.8 & 103.4 & 12.8 & 26.3 & 39 & 15 \\ \hline \hline
ssFix & 0.6 & 118.1 & 8.5 & 23.9 & 43 & 11 \\ \hline
jGenProg & 1.8 & 61.9 & 14.6 & 20.7 & 5 & 1 \\ \hline
jKali & 1.2 & 32.7 & 21.6 & 18.8 & 6 & 1 \\ \hline
Nopol & 4.3 & 29 & 9.5 & 12.6 & 8 & 0 \\ \hline
HDRepair & 93.8 & 108.1 & 101 & 101 & 2 & 1 \\ \hline
\end{tabular}
\end{threeparttable}
\vspace{-3em}
\end{table}

\Cref{bugs_dot_jar_eval_table0} shows the repairing results of sharpFix
and ssFix for each of the 7 projects and for all of them. sharpFix has 
better repair performance than ssFix:
it produced four more correct patches and was less prone to producing
overfitting (plausible-but-incorrect) patches with the non-overfitting
rate being 15/39=38.5\%. Though sharpFix ran longer than ssFix did to 
produce a plausible patch, the running times are still comparable. For 
the five projects: ACC, CML, FLK, OAK, and MNG, our results show that 
sharpFix and ssFix are comparable but both have limited repair abilities. 
For MAT, sharpFix
produced an additional correct patch but ran slower than ssFix. For WCT,
sharpFix outperformed ssFix in correctly repairing 6 bugs that ssFix did not.
We found that for five of the bugs: WCT-3098, WCT-3520, WCT-3845, WCT-4276, 
and WCT-5891, sharpFix found very effective candidates: 
for each such bug, it looked at no more than 8 candidates to yield the 
correct patch. For WCT-5686, sharpFix reused a candidate ranked 31th to
produce the correct patch. ssFix failed to reuse any candidates to produce
any plausible patches for four bugs: WCT-3098, WCT-3520, WCT-3845, and
WCT-5686. For the other two bugs: WCT-4276 and WCT-5891, it only produced
overfitting patches.

\Cref{bugs_dot_jar_eval_table1} shows the repairing results of all the
six techniques. We found that compared to sharpFix and ssFix, the other
four techniques have very limited repair abilities. They each can only
produce correct patches for no more than one bug. jGenProg only looks 
at finding the fix ingredients as statements from the local faulty program. 
This type of repair constrains itself from finding useful fix ingredients that
are expressions and are from non-local programs. jKali can only do deletions
and is unable to produce many types of non-deletion patches. Nopol looks at 
producing if-condition-related patches but is prone to synthesizing if-conditions 
that are either too constrained or too loose. HDRepair leverages mined
bug-fixing changes to guide the search of a correct patch. However, according 
to our results, this type of guidance is not effective.

All the fix ingredients, retrieved candidates, and repair results are released
and can be found at \textit{https://github.com/sharpFix18/sharpFix/tree/master/expt0}.
\vspace{-1.5em}

\subsection{Discussion}

For all repair experiments, we determined the correctness of a plausible patch
by manually comparing it to the developer patch available from the dataset
(either Defects4J or Bugs.jar-ELIXIR). We took a relatively conservative approach
for our manual comparing process and only considered a patch as correct if we 
found a semantics-preserving transformation between the generated patch and the
developer patch. Since our manual process is conservative, the correctness of
the patches that we identified as correct are clear. We released all such patches
and provided reasons for why they are correct.

ELIXIR was not available when we performed our experiments, and we did not run 
it on the Bugs.jar-ELIXIR dataset for comparison.
In \cite{saha17elixir}, ELIXIR was shown to repair 22 bugs with correct 
patches generated (the times for generating these patches however were not given).
For repair, ELIXIR leveraged bug reports for ranking the generated patches, and
this was shown to be very effective: for the Defects4J bugs, ELIXIR generated 
6/(26-6)=30\% more patches using bug reports than it did without using bug reports. 
sharpFix did not leverage any bug report information for repair: such information
is typically not available. Though sharpFix
outperformed ELIXIR in generating 10 more correct patches for the Defects4J bugs,
it generated 7 fewer correct patches for the Bugs.jar-ELIXIR bugs. It would be
interesting to see how sharpFix works by leveraging the bug report information. 
It would also be interesting to compare the two techniques on other possible 
datasets. We leave these as future work.

\section{Related Work} \label{related_work}

In this paper, we conducted an experiment testing the fix ingredient
assumption. Our experiment is closely related to the studies by 
Barr et al. \cite{barr14} and by Sumi et al. \cite{sumi15} which also 
investigate whether the fix ingredients exist in local and non-local programs.
Different from our experiment, the two studies looked at the code line(s) 
as a fix ingredient and investigated what fraction of a fix ingredient
(in terms of the code lines) for a bug-fixing change can be found from
existing programs. Our experiment is also related to other studies
that look at the repetitiveness of a fixing change (rather than the fix code)
\cite{nguyen13study}, the existence of fix ingredients in the program's
earlier versions \cite{martinez14}, and code redundancy
\cite{gabel10,hindle12,lin17uniqueness} in general.

sharpFix finds and reuses existing code fragments from the local faulty
program and a code repository to do bug repair, and is an improved version
of ssFix \cite{xin17leveraging}. sharpFix is closely related to
SearchRepair \cite{ke15} and CodePhage \cite{douskos15} which also do code
search to find existing code for bug repair. Different from sharpFix which
performs syntactic code search, SearchRepair's code search is based on
symbolic execution and constraint-solving, and CodePhage's code search is based
on program execution. CSAR \cite{hill18automated} is an improvement over 
SearchRepair. It performs string matching on constraints rather than
doing constraint-solving to identify semantics-related code. sharpFix is
also related to the recent technique SimFix \cite{jiang18shaping} which
leverages similar code to produce patches.
Different from sharpFix, SimFix also leverages existing patches to build
the search space and it only looks at the local program for finding similar code. 
The syntactic features that SimFix and sharpFix use for
finding similar code are also different. GenProg \cite{legoues12,legoues12-icse}
is an early APR technique that reuses existing code to produce patches and is
related to sharpFix. Different from sharpFix, GenProg only reuses statements
from the local faulty program itself to produce patches. It does not directly
find and reuse similar code but uses genetic algorithms to modify the program
and produce patches.

sharpFix is related to many repair techniques that use
genetic algorithms \cite{legoues12,legoues12-icse,weimer13}, random
search \cite{y_qi14}, human-written templates \cite{kim13}, bug-fixing
instances \cite{gao15,le16,liu18mining}, program comparison \cite{tan15}, 
program synthesis \cite{nguyen13,mechtaev15,mechtaev16,dantoni16,le17s3,singh13},
condition synthesis \cite{xuan16,xiong17}, repair templates plus condition
synthesis \cite{long15}, modifications with patch ranking models 
\cite{long16,saha17elixir,wen18context}, modifications based on monitored program 
states \cite{chen17contract}, learned transformations \cite{rolim17,long17},
invariants \cite{perkins09automatically}, bug reports \cite{liu13r2fix},
statistical analysis \cite{kaleeswaran14},
reference implementation \cite{mechtaev18semantic}, and
non-test-suite specifications \cite{wei10,gopinath11,van18static}.

\section{Conclusion}

In this paper, we conducted multiple experiments for analyzing
ssFix, a syntactic search-based repair technique. We found
the built-upon idea of ssFix (and other related techniques), i.e.,
reusing existing code for bug repair is promising. However, the 
approaches used by ssFix for code search and code reuse can be
significantly improved. We developed a new repair technique sharpFix
which uses improved approaches for code search and code reuse
and demonstrated through experiments that it outperforms ssFix and 
four other repair techniques.

\balance
\bibliographystyle{IEEEtran}
\bibliography{fix_reuse}

\begin{thebibliography}{10}
\providecommand{\url}[1]{#1}
\csname url@samestyle\endcsname
\providecommand{\newblock}{\relax}
\providecommand{\bibinfo}[2]{#2}
\providecommand{\BIBentrySTDinterwordspacing}{\spaceskip=0pt\relax}
\providecommand{\BIBentryALTinterwordstretchfactor}{4}
\providecommand{\BIBentryALTinterwordspacing}{\spaceskip=\fontdimen2\font plus
\BIBentryALTinterwordstretchfactor\fontdimen3\font minus
  \fontdimen4\font\relax}
\providecommand{\BIBforeignlanguage}[2]{{%
\expandafter\ifx\csname l@#1\endcsname\relax
\typeout{** WARNING: IEEEtran.bst: No hyphenation pattern has been}%
\typeout{** loaded for the language `#1'. Using the pattern for}%
\typeout{** the default language instead.}%
\else
\language=\csname l@#1\endcsname
\fi
#2}}
\providecommand{\BIBdecl}{\relax}
\BIBdecl

\bibitem{monperrus18automatic}
M.~Monperrus, ``Automatic software repair: a bibliography,'' \emph{ACM
  Computing Surveys (CSUR)}, vol.~51, no.~1, p.~17, 2018.

\bibitem{legoues12}
C.~L. Goues, T.~Nguyen, S.~Forrest, and W.~Weimer, ``{GenProg}: A generic
  method for automatic software repair,'' \emph{IEEE Transactions on Software
  Engineering (TSE)}, vol.~38, no.~1, pp. 54--72, 2012.

\bibitem{weimer13}
W.~Weimer, Z.~P. Fry, and S.~Forrest, ``Leveraging program equivalence for
  adaptive program repair: models and first results,'' in \emph{Proceedings of
  the 28th International Conference on Automated Software Engineering
  (ASE)}.\hskip 1em plus 0.5em minus 0.4em\relax IEEE, 2013, pp. 356--366.

\bibitem{kim13}
D.~Kim, J.~Nam, J.~Song, and S.~Kim, ``Automatic patch generation learned from
  human-written patches,'' in \emph{Proceedings of the 2013 International
  Conference on Software Engineering (ICSE)}.\hskip 1em plus 0.5em minus
  0.4em\relax IEEE Press, 2013, pp. 802--811.

\bibitem{y_qi14}
Y.~Qi, X.~Mao, Y.~Lei, Z.~Dai, and C.~Wang, ``The strength of random search on
  automated program repair,'' in \emph{Proceedings of the 36th International
  Conference on Software Engineering (ICSE)}.\hskip 1em plus 0.5em minus
  0.4em\relax ACM, 2014, pp. 254--265.

\bibitem{ke15}
Y.~Ke, K.~T. Stolee, C.~Le~Goues, and Y.~Brun, ``Repairing programs with
  semantic code search (t),'' in \emph{Proceedings of 30th IEEE/ACM
  International Conference on Automated Software Engineering (ASE)}.\hskip 1em
  plus 0.5em minus 0.4em\relax IEEE, 2015, pp. 295--306.

\bibitem{z_qi15}
Z.~Qi, F.~Long, S.~Achour, and M.~Rinard, ``An analysis of patch plausibility
  and correctness for generate-and-validate patch generation systems,'' in
  \emph{Proceedings of the 2015 International Symposium on Software Testing and
  Analysis (ISSTA)}.\hskip 1em plus 0.5em minus 0.4em\relax ACM, 2015, pp.
  24--36.

\bibitem{xuan16}
\BIBentryALTinterwordspacing
J.~Xuan, M.~Martinez, F.~DeMarco, M.~Cl{\'e}ment, S.~Lamelas, T.~Durieux,
  D.~Le~Berre, and M.~Monperrus, ``{Nopol: Automatic Repair of Conditional
  Statement Bugs in Java Programs},'' \emph{IEEE Transactions on Software
  Engineering (TSE)}, vol.~43, pp. 34--55, 2016. [Online]. Available:
  \url{https://hal.archives-ouvertes.fr/hal-01285008/document}
\BIBentrySTDinterwordspacing

\bibitem{le16}
X.~B.~D. Le, D.~Lo, and C.~Le~Goues, ``History driven program repair,'' in
  \emph{Proceedings of the 23rd International Conference on Software Analysis,
  Evolution, and Reengineering (SANER)}, vol.~1.\hskip 1em plus 0.5em minus
  0.4em\relax IEEE, 2016, pp. 213--224.

\bibitem{long16}
F.~Long and M.~Rinard, ``Automatic patch generation by learning correct code,''
  in \emph{In Proceedings of the 43rd Annual ACM SIGPLAN-SIGACT Symposium on
  Principles of Programming Languages (POPL)}.\hskip 1em plus 0.5em minus
  0.4em\relax ACM, 2016, pp. 298--312.

\bibitem{mechtaev16}
S.~Mechtaev, J.~Yi, and A.~Roychoudhury, ``Angelix: Scalable multiline program
  patch synthesis via symbolic analysis,'' in \emph{Proceedings of the 38th
  International Conference on Software Engineering (ICSE)}.\hskip 1em plus
  0.5em minus 0.4em\relax IEEE, 2016, pp. 691--701.

\bibitem{le17s3}
X.-B.~D. Le, D.-H. Chu, D.~Lo, C.~Le~Goues, and W.~Visser, ``{S3}: syntax- and
  semantic-guided repair synthesis via programming by examples,'' in
  \emph{Proceedings of the 11th Joint Meeting on Foundations of Software
  Engineering (ESEC/FSE)}.\hskip 1em plus 0.5em minus 0.4em\relax ACM, 2017,
  pp. 593--604.

\bibitem{xiong17}
Y.~Xiong, J.~Wang, R.~Yan, J.~Zhang, S.~Han, G.~Huang, and L.~Zhang, ``Precise
  condition synthesis for program repair,'' in \emph{Proceedings of the 39th
  International Conference on Software Engineering (ICSE)}.\hskip 1em plus
  0.5em minus 0.4em\relax IEEE, 2017, pp. 416--426.

\bibitem{chen17contract}
L.~Chen, Y.~Pei, and C.~A. Furia, ``Contract-based program repair without the
  contracts,'' in \emph{Proceedings of the 32nd IEEE/ACM International
  Conference on Automated Software Engineering (ASE)}.\hskip 1em plus 0.5em
  minus 0.4em\relax IEEE, 2017, pp. 637--647.

\bibitem{saha17elixir}
R.~K. Saha, Y.~Lyu, H.~Yoshida, and M.~R. Prasad, ``Elixir: effective object
  oriented program repair,'' in \emph{Proceedings of the 32nd IEEE/ACM
  International Conference on Automated Software Engineering (ASE)}.\hskip 1em
  plus 0.5em minus 0.4em\relax IEEE, 2017, pp. 648--659.

\bibitem{xin17leveraging}
Q.~Xin and S.~P. Reiss, ``Leveraging syntax-related code for automated program
  repair,'' in \emph{Proceedings of the 32nd IEEE/ACM International Conference
  on Automated Software Engineering (ASE)}.\hskip 1em plus 0.5em minus
  0.4em\relax IEEE, 2017, pp. 660--670.

\bibitem{liu18mining}
X.~Liu and H.~Zhong, ``Mining stackoverflow for program repair,'' in
  \emph{Proceedings of the 25th International Conference on Software Analysis,
  Evolution and Reengineering (SANER)}.\hskip 1em plus 0.5em minus 0.4em\relax
  IEEE, 2018, pp. 118--129.

\bibitem{jiang18shaping}
J.~Jiang, Y.~Xiong, H.~Zhang, Q.~Gao, and X.~Chen, ``Shaping program repair
  space with existing patches and similar code,'' in \emph{Proceedings of the
  27th International Symposium on Software Testing and Analysis (ISSTA)}.\hskip
  1em plus 0.5em minus 0.4em\relax ACM, 2018, pp. 298--309.

\bibitem{hua18towards}
J.~Hua, M.~Zhang, K.~Wang, and S.~Khurshid, ``Towards practical program repair
  with on-demand candidate generation,'' in \emph{Proceedings of the 40th
  International Conference on Software Engineering (ICSE)}.\hskip 1em plus
  0.5em minus 0.4em\relax ACM, 2018, pp. 12--23.

\bibitem{mechtaev18semantic}
S.~Mechtaev, M.-D. Nguyen, Y.~Noller, L.~Grunske, and A.~Roychoudhury,
  ``Semantic program repair using a reference implementation,'' in
  \emph{Proceedings of the 40th International Conference on Software
  Engineering (ICSE)}.\hskip 1em plus 0.5em minus 0.4em\relax ACM, 2018.

\bibitem{wen18context}
M.~Wen, J.~Chen, R.~Wu, D.~Hao, and S.-C. Cheung, ``Context-aware patch
  generation for better automated program repair,'' in \emph{Proceedings of the
  40th International Conference on Software Engineering (ICSE)}.\hskip 1em plus
  0.5em minus 0.4em\relax ACM, 2018.

\bibitem{long16-anal}
F.~Long and M.~Rinard, ``An analysis of the search spaces for generate and
  validate patch generation systems,'' in \emph{Proceedings of the 38th
  International Conference on Software Engineering (ICSE)}.\hskip 1em plus
  0.5em minus 0.4em\relax ACM, 2016, pp. 702--713.

\bibitem{defects4j}
R.~Just, D.~Jalali, and M.~D. Ernst, ``{Defects4J}: A database of existing
  faults to enable controlled testing studies for {Java} programs,'' in
  \emph{Proceedings of the 2014 International Symposium on Software Testing and
  Analysis (ESEC/FSE)}.\hskip 1em plus 0.5em minus 0.4em\relax ACM, 2014, pp.
  437--440.

\bibitem{astorimpl}
``{SpoonLabs Astor},'' https://github.com/SpoonLabs/astor.

\bibitem{nopolimpl}
``{SpoonLabs Nopol},'' https://github.com/SpoonLabs/nopol.

\bibitem{hdrepairimpl}
{HDRepair}, ``{HDRepair} repository,'' https://github.com/xuanbachle/bugfixes,
  2016.

\bibitem{barr14}
E.~T. Barr, Y.~Brun, P.~Devanbu, M.~Harman, and F.~Sarro, ``The plastic surgery
  hypothesis,'' in \emph{Proceedings of the 22nd ACM SIGSOFT International
  Symposium on Foundations of Software Engineering (ESEC/FSE)}.\hskip 1em plus
  0.5em minus 0.4em\relax ACM, 2014, pp. 306--317.

\bibitem{sumi15}
S.~Sumi, Y.~Higo, K.~Hotta, and S.~Kusumoto, ``Toward improving graftability on
  automated program repair,'' in \emph{Proceedings of the 2015 IEEE
  International Conference on Software Maintenance and Evolution
  (ICSME)}.\hskip 1em plus 0.5em minus 0.4em\relax IEEE, 2015, pp. 511--515.

\bibitem{nguyen13study}
H.~A. Nguyen, A.~T. Nguyen, T.~T. Nguyen, T.~N. Nguyen, and H.~Rajan, ``A study
  of repetitiveness of code changes in software evolution,'' in
  \emph{Proceedings of the 28th International Conference on Automated Software
  Engineering (ASE)}.\hskip 1em plus 0.5em minus 0.4em\relax IEEE, 2013, pp.
  180--190.

\bibitem{martinez14}
M.~Martinez, W.~Weimer, and M.~Monperrus, ``Do the fix ingredients already
  exist? an empirical inquiry into the redundancy assumptions of program repair
  approaches,'' in \emph{Companion Proceedings of the 36th International
  Conference on Software Engineering (ICSE)}.\hskip 1em plus 0.5em minus
  0.4em\relax ACM, 2014, pp. 492--495.

\bibitem{long15}
F.~Long and M.~Rinard, ``Staged program repair with condition synthesis,'' in
  \emph{Proceedings of the 10th Joint Meeting on Foundations of Software
  Engineering (ESEC/FSE)}.\hskip 1em plus 0.5em minus 0.4em\relax ACM, 2015,
  pp. 166--178.

\bibitem{darpamuse}
{DARPA MUSE}, ``{DARPA MUSE} repository,''
  https://opencatalog.darpa.mil/MUSE.html, 2016.

\bibitem{campos12}
J.~Campos, A.~Riboira, A.~Perez, and R.~Abreu, ``{GZoltar}: an eclipse plug-in
  for testing and debugging,'' in \emph{Proceedings of the 27th IEEE/ACM
  International Conference on Automated Software Engineering (ASE)}.\hskip 1em
  plus 0.5em minus 0.4em\relax ACM, 2012, pp. 378--381.

\bibitem{reiss09semantics}
S.~P. Reiss, ``Semantics-based code search,'' in \emph{Proceedings of the 31st
  International Conference on Software Engineering (ICSE)}.\hskip 1em plus
  0.5em minus 0.4em\relax IEEE, 2009, pp. 243--253.

\bibitem{cornu15npefix}
\BIBentryALTinterwordspacing
B.~Cornu, T.~Durieux, L.~Seinturier, and M.~Monperrus, ``{NPEFix: Automatic
  Runtime Repair of Null Pointer Exceptions in Java},'' 2015, working paper or
  preprint. [Online]. Available:
  \url{https://hal.archives-ouvertes.fr/hal-01251960}
\BIBentrySTDinterwordspacing

\bibitem{eclipsejdt}
``{Eclipse JDT},'' \url{https://www.eclipse.org/jdt}.

\bibitem{saha18bdj}
R.~K. Saha, Y.~Lyu, W.~Lam, H.~Yoshida, and M.~R. Prasad, ``{Bugs.jar}: a
  large-scale, diverse dataset of real-world java bugs,'' in \emph{Proceedings
  of the 15th International Conference on Mining Software Repositories
  (MSR)}.\hskip 1em plus 0.5em minus 0.4em\relax ACM, 2018, pp. 10--13.

\bibitem{douskos15}
S.~Sidiroglou-Douskos, E.~Lahtinen, F.~Long, and M.~Rinard, ``Automatic error
  elimination by horizontal code transfer across multiple applications,'' in
  \emph{Proceedings of the 36th ACM SIGPLAN Conference on Programming Language
  Design and Implementation (PLDI)}.\hskip 1em plus 0.5em minus 0.4em\relax
  ACM, 2015, pp. 43--54.

\bibitem{gabel10}
M.~Gabel and Z.~Su, ``A study of the uniqueness of source code,'' in
  \emph{Proceedings of the eighteenth ACM SIGSOFT international symposium on
  Foundations of software engineering (ESEC/FSE)}.\hskip 1em plus 0.5em minus
  0.4em\relax ACM, 2010, pp. 147--156.

\bibitem{hindle12}
A.~Hindle, E.~T. Barr, Z.~Su, M.~Gabel, and P.~Devanbu, ``On the naturalness of
  software,'' in \emph{Proceedings of the 34th International Conference on
  Software Engineering (ICSE)}.\hskip 1em plus 0.5em minus 0.4em\relax IEEE,
  2012, pp. 837--847.

\bibitem{lin17uniqueness}
B.~Lin, L.~Ponzanelli, A.~Mocci, G.~Bavota, and M.~Lanza, ``On the uniqueness
  of code redundancies,'' in \emph{Proceedings of the 25th International
  Conference on Program Comprehension (ICPC)}.\hskip 1em plus 0.5em minus
  0.4em\relax IEEE, 2017, pp. 121--131.

\bibitem{hill18automated}
A.~Hill, C.~S. P{\u{a}}s{\u{a}}reanu, and K.~T. Stolee, ``Automated program
  repair with canonical constraints,'' in \emph{Proceedings of the 40th
  International Conference on Software Engineering (ICSE): Companion
  Proceeedings}.\hskip 1em plus 0.5em minus 0.4em\relax ACM, 2018, pp.
  339--341.

\bibitem{legoues12-icse}
C.~L. Goues, M.~Dewey-Vogt, S.~Forrest, and W.~Weimer, ``A systematic study of
  automated program repair: fixing 55 out of 105 bugs for \$8 each,'' in
  \emph{Proceedings of the 34th International Conference on Software
  Engineering (ICSE)}.\hskip 1em plus 0.5em minus 0.4em\relax IEEE, 2012, pp.
  3--13.

\bibitem{gao15}
Q.~Gao, H.~Zhang, J.~Wang, Y.~Xiong, L.~Zhang, and H.~Mei, ``Fixing recurring
  crash bugs via analyzing q\&a sites (t),'' in \emph{Proceedings of the 30th
  IEEE/ACM International Conference on Automated Software Engineering
  (ASE)}.\hskip 1em plus 0.5em minus 0.4em\relax IEEE, 2015, pp. 307--318.

\bibitem{tan15}
S.~H. Tan and A.~Roychoudhury, ``{relifix}: Automated repair of software
  regressions,'' in \emph{Proceedings of the 37th International Conference on
  Software Engineering (ICSE)}.\hskip 1em plus 0.5em minus 0.4em\relax IEEE,
  2015, pp. 471--482.

\bibitem{nguyen13}
H.~D.~T. Nguyen, D.~Qi, A.~Roychoudhury, and S.~Chandra, ``{SemFix}: Program
  repair via semantic analysis,'' in \emph{Proceedings of the 2013
  International Conference on Software Engineering (ICSE)}.\hskip 1em plus
  0.5em minus 0.4em\relax IEEE, 2013, pp. 772--781.

\bibitem{mechtaev15}
S.~Mechtaev, J.~Yi, and A.~Roychoudhury, ``{DirectFix}: Looking for simple
  program repairs,'' in \emph{Proceedings of the 37th International Conference
  on Software Engineering (ICSE)}.\hskip 1em plus 0.5em minus 0.4em\relax IEEE,
  2015, pp. 448--458.

\bibitem{dantoni16}
L.~D'Antoni, R.~Samanta, and R.~Singh, ``Qlose: Program repair with
  quantitative objectives,'' in \emph{International Conference on Computer
  Aided Verification (CAV)}.\hskip 1em plus 0.5em minus 0.4em\relax Springer,
  2016, pp. 383--401.

\bibitem{singh13}
R.~Singh, S.~Gulwani, and A.~Solar-Lezama, ``Automated feedback generation for
  introductory programming assignments,'' in \emph{PLDI}, 2013, pp. 15--26.

\bibitem{rolim17}
R.~Rolim, G.~Soares, D.~Loris, O.~Polozov, S.~Gulwani, R.~Gheyi, R.~Suzuki, and
  B.~Hartmann, ``Learning syntactic program transformations from examples,'' in
  \emph{Proceedings of the 39th International Conference on Software
  Engineering (ICSE)}.\hskip 1em plus 0.5em minus 0.4em\relax IEEE, 2017, pp.
  404--415.

\bibitem{long17}
F.~Long, P.~Amidon, and M.~Rinard, ``Automatic inference of code transforms for
  patch generation,'' in \emph{Proceedings of the 11th Joint Meeting on
  Foundations of Software Engineering (ESEC/FSE)}.\hskip 1em plus 0.5em minus
  0.4em\relax ACM, 2017, pp. 727--739.

\bibitem{perkins09automatically}
J.~H. Perkins, S.~Kim, S.~Larsen, S.~Amarasinghe, J.~Bachrach, M.~Carbin,
  C.~Pacheco, F.~Sherwood, S.~Sidiroglou, G.~Sullivan \emph{et~al.},
  ``Automatically patching errors in deployed software,'' in \emph{Proceedings
  of the ACM SIGOPS 22nd symposium on Operating systems principles
  (SOSP)}.\hskip 1em plus 0.5em minus 0.4em\relax ACM, 2009, pp. 87--102.

\bibitem{liu13r2fix}
C.~Liu, J.~Yang, L.~Tan, and M.~Hafiz, ``{R2Fix}: Automatically generating bug
  fixes from bug reports,'' in \emph{Proceedings of the Sixth International
  Conference on Software Testing, Verification and Validation (ICST)}.\hskip
  1em plus 0.5em minus 0.4em\relax IEEE, 2013, pp. 282--291.

\bibitem{kaleeswaran14}
S.~Kaleeswaran, V.~Tulsian, A.~Kanade, and A.~Orso, ``{MintHint}: Automated
  synthesis of repair hints,'' in \emph{Proceedings of the 36th International
  Conference on Software Engineering (ICSE)}.\hskip 1em plus 0.5em minus
  0.4em\relax ACM, 2014, pp. 266--276.

\bibitem{wei10}
Y.~Wei, Y.~Pei, C.~A. Furia, S.~L. S, S.~Buchholz, M.~B., and A.~Zeller,
  ``Automated fixing of programs with contracts,'' in \emph{Proceedings of the
  19th international symposium on Software testing and analysis (ISSTA)}.\hskip
  1em plus 0.5em minus 0.4em\relax ACM, 2010, pp. 61--72.

\bibitem{gopinath11}
D.~Gopinath, M.~Z. Malik, and S.~Khurshid, ``Specification-based program repair
  using {SAT},'' in \emph{Proceedings of the 17th international conference on
  Tools and algorithms for the construction and analysis of systems
  (TACAS)}.\hskip 1em plus 0.5em minus 0.4em\relax Springer, 2011, pp.
  173--188.

\bibitem{van18static}
R.~van Tonder and C.~Le~Goues, ``Static automated program repair for heap
  properties,'' in \emph{Proceedings of the 40th International Conference on
  Software Engineering (ICSE)}.\hskip 1em plus 0.5em minus 0.4em\relax ACM,
  2018, pp. 151--162.

\end{thebibliography}
\end{document}